\documentclass[conference, a4paper, svgnames]{IEEEtran}
\IEEEoverridecommandlockouts
\usepackage{cite}
\usepackage{amsmath,amssymb,amsfonts}
\usepackage{algorithmic}
\usepackage{graphicx}
\usepackage{textcomp}
\usepackage{xcolor}
\usepackage{multirow}
\usepackage{booktabs}
\usepackage{bbold}
\usepackage{url}
\usepackage{siunitx}
\usepackage{subcaption}
\usepackage{cuted}
\usepackage[caption=false]{subfig}
\usepackage[nolist]{acronym}
\usepackage[inline]{enumitem}
\usepackage{optidef}
\usepackage{fnpos}
\renewcommand{\baselinestretch}{0.863}
\usepackage[margin=0.72in]{geometry}       % 0.77, seems to be IEEE setting
\begin{document}

\makeatletter
%%%%%%%%%%%%%%%%%%%%%%%%%%%%%% User specified LaTeX commands.

\title{Uncertainty-Aware Flexibility of Buildings: \\ From Quantification to Provision 
\thanks{This research is supported by NCCR Automation, a National Centre of Competence in Research, funded by the SNSF (grant number 180545).}
}

\author{\IEEEauthorblockN{Julie Rousseau\IEEEauthorrefmark{1}\IEEEauthorrefmark{3}\IEEEauthorrefmark{2},
Hanmin Cai\IEEEauthorrefmark{3},
Philipp Heer\IEEEauthorrefmark{3},
Kristina Orehounig\IEEEauthorrefmark{4},
Gabriela Hug\IEEEauthorrefmark{1}}
\IEEEauthorblockA{\IEEEauthorrefmark{1}Power Systems Laboratory, ETH Zürich, Zürich, Switzerland}
\IEEEauthorblockA{\IEEEauthorrefmark{3} Urban Energy Systems Laboratory, Empa, D\"{u}bendorf, Switzerland}
\IEEEauthorblockA{\IEEEauthorrefmark{4} Research unit for Building Physics and Building Ecology, Vienna University of Technology, Vienna, Austria}
\IEEEauthorblockA{\IEEEauthorrefmark{2}jrousseau@ethz.ch}
}

\maketitle

\begin{abstract}

Buildings represent a promising flexibility source to support the integration of renewable energy sources, as they may shift their heating energy consumption over time without impacting users’ comfort. However, a building's predicted flexibility potential is based on uncertain ambient weather forecasts and a typically inaccurate building thermal model. Hence, this paper presents an uncertainty-aware flexibility quantifier using a chance-constrained formulation. Because such a quantifier may be conservative, we additionally model real-time feedback in the quantification, in the form of affine feedback policies. Such adaptation can take the form of intra-day trades or rebound around the flexibility provision period. To assess the flexibility quantification formulations, we further assume that flexible buildings participate in secondary frequency control markets. The results show some increase in flexibility and revenues when introducing affine feedback policies. Additionally, it is demonstrated that accounting for uncertainties in the flexibility quantification is necessary, especially when intra-day trades are not available. Even though an uncertainty-ignorant potential may seem financially profitable in secondary frequency control markets, it comes at the cost of significant thermal discomfort for inhabitants. Hence, we suggest a comfort-preserving approach, aiming to truly reflect thermal discomfort on the economic flexibility revenue, to obtain a fairer comparison.  
\end{abstract}


\begin{acronym}[ML] 
\acro{MFPH}{Maximum Flexibility Provision Horizon}
\acro{UMAR}{Urban Mining And Recycling}
\acro{aFRR}{automatic Frequency Restoration Reserve}
\acro{TSO}{Transmission System Operator}
\acro{BRP}{Balancing Responsible Party}
\acro{BSP}{Balancing Service Provider}
\acro{FEA}{Flexibility Envelope Area}
\end{acronym}

\section{Introduction}

In an attempt to reduce global carbon emissions, power systems undergo profound transformations: new renewable power plants are installed, while a significant share of conventional units are about to be decommissioned \cite{ReportIRENA2019}. These changes call for additional and new sources of flexibility to balance the daily intermittency and variability brought by non-dispatchable renewable power plants, as well as to compensate for the loss of flexibility provided by conventional generators \cite{EUFlexibility}. 

Buildings' power consumption can be flexible, e.g., by shifting their heating power consumption over time while preserving the inhabitants' thermal comfort \cite{IEAFlexibilityQuantification}. Besides, buildings' electricity consumption represents a large share of the total consumption. In Switzerland, in 2021, 33.9\% of the electricity consumption was used to supply buildings, out of which 25\% was used for heating systems \cite{BFEstatistics2021}, a number expected to increase in the coming years \cite{EnergyStrategy2050}. Hence, buildings do not only have the potential to be flexible, but their overall potential, at a national scale, is large.

The technical and economic feasibility of flexible building operations has been extensively investigated. 
Based on their technical characteristics, buildings can participate in a variety of existing flexibility markets, including frequency control markets characterized by short-time responses. Their technical performances have been tested in detailed simulation frameworks \cite{ZHAO2013}, as well as in existing controllable facilities \cite{vrettos2018}. Besides, by selling their flexibility as a service, buildings may earn revenues. Even though the economic value of a buiding's flexibility remains small under the current conditions, it still brings savings to households \cite{LAMPROPOULOS2019100187, ZHOU2022e09865}.

\subsection{Building's Flexibility Quantification}
To participate in flexibility markets, buildings must quantify their flexibility, i.e., their feasible electricity consumption region. The power rating of a building's heating system limits the flexible power, and the inhabitant's thermal comfort requirements introduce energy constraints. No consensus exists in the literature on flexibility metrics of energy-constrained systems. Power trajectories spanning the feasible power consumption paths can describe the flexibility of such systems, at the cost of significant computation efforts \cite{PINTO2017}. The deviation from a baseline power consumption in response to a penalty signal also describes a building's flexibility but fails to capture time-coupling energy constraints \cite{junker2018characterizing, Oldewurtel2013}. Energy flexibility envelopes on the other hand represent a promising metric for energy-constrained systems, as they describe a building's feasible energy consumption region, delimited by the smallest and largest energy amount that can be absorbed without violating technical or comfort constraints \cite{Cai2021, DECONINCK2016, Rousseau2023}. They also offer a standardized representation of heterogeneous energy-constrained systems' flexibility and are, therefore, employed in this paper. 

Tractable and efficient flexibility computations often require a simplified representation of a building's thermal dynamics \cite{KATHIRGAMANATHAN2021110120}. Resistance-capacitance linear models provide a simple building model but disregard non-linear thermal dynamics \cite{FINCK2019}. However, the accuracy of such low-order models significantly decreases as the prediction horizon increases \cite{FONTI2017}, yielding an unreliable flexibility estimation for timesteps further in the future. When more complex building models are used, flexibility characterization tends to be oversimplified \cite{LIND2023113698} or computationally expensive \cite{manolisThesis}. Therefore, in this paper, we aim to extend the methodology introduced in \cite{Rousseau2023} that employs a low-order building thermal model but acknowledges the reduced model accuracy over future horizons.

\subsection{Building's Flexibility under Uncertainties}

The impact of uncertainties on an energy-constrained system is particularly significant as uncertainties accumulate over time, yielding a poor estimation of the future system's state. Some studies have investigated the impact of accounting for uncertainties on the future flexibility of a building's heating system. The authors of \cite{Vrakopoulou2019} investigate the flexibility potential of buildings' heating systems when describing building thermal dynamics with a low-order model but only assume ambient weather uncertainties, neglecting the model inaccuracy. Similarly, the authors in \cite{AMADEH2022} quantify the flexibility of a residential building, represented as a low-order linear model, using energy envelopes but only consider uncertainties in ambient weather and initial indoor conditions. The authors in \cite{SCHARNHORST2022} compute the energy envelopes of a residential building and consider the modeling inaccuracy of a low-order thermal dynamics model, but test their methodology on a single-zone residential building, simulated with a low-order model, and therefore obtain an unrealistically high modeling accuracy. 

In \cite{Rousseau2023}, we study the impact of accounting for ambient weather prediction and thermal dynamics modeling inaccuracy on the flexibility potential of a residential building's heating system. We propose an uncertainty-aware energy envelope formulation that ensures the satisfaction of constraints with a fixed confidence level using chance-constrained optimization. Based on a real occupied residential building, we demonstrate a dominant impact of modeling inaccuracy, significantly restricting the flexibility of a residential building. This impact mostly results from stochastic inhabitants' behaviour, neglected in low-order building models.  

\subsection{Introducing Feedback}
\label{sec:introFeedback}

Chance-constrained optimizations may result in conservative optimal solutions which are robust to the accumulation of uncertainties over the optimization horizon, i.e., the system is assumed to operate in an open loop. However, in closed-loop operation, systems constantly adapt to new conditions. Modeling this closed-loop feedback in chance-constrained optimizations may yield less conservative solutions \cite{GOULART2006}. However, representing feedback in stochastic optimization results, in general, in an intractable formulation. Nevertheless, some particular feedback policies, e.g., a linear adaptation of the system to past states over a finite horizon, referred to as affine feedback policy, result in tractable chance-constrained optimization \cite{BenTak2004, OldewurtelTractableCC}.

Quantifying energy flexibility bounds can be considered an open or closed-loop problem, depending on the application. For instance, integrating the energy bounds into a receding horizon dispatch optimization with regular updates of the bounds describes a closed-loop set-up. Alternatively, using flexibility envelopes to dispatch an entire horizon's flexibility constitutes an open-loop flexibility usage. In this work, we assume an open-loop flexibility set-up that is further described in Section~\ref{sec:bidding&clearing}. 

However, feedback may also exist in an open-loop flexibility setup. Indeed, when defining flexibility as a deviation from a baseline power consumption, previous studies suggest constantly adjusting resources' baseline consumption while delivering the promised flexibility \cite{BUNNING2022, GORECKI2017229}. Hence, based on the latest measured indoor temperature, the resource power baseline consumption may be adapted, while the deviation from this baseline equals the promised value. 

Different set-ups are conceivable in power baseline adaptation. 
Some studies suggest to constantly adapt a flexible system's baseline through intra-day trades. In \cite{BUNNING2022}, a flexible heat pump coupled with thermal storage participates in primary frequency control and is allowed to trade its baseline power consumption changes in intra-day markets. The authors report a significant revenue increase when introducing real-time feedback in the flexibility quantification. In \cite{GORECKI2017229}, a flexible heat pump provides secondary frequency control, considering only uncertain future flexibility requests. By implementing the method in an existing building, it is concluded that, with intra-day baseline adaptation, the flexibility revenues and the inhabitants' thermal comfort can increase. As an alternative to intra-day markets, the collocation of a controllable power source allows the adaptation of a flexible resource's baseline \cite{li2023unlocking}.

Even though such adaptations are theoretically possible, they may be challenging to implement in practice. Indeed, in Continental Europe, two actors intervene in the energy management of flexible resources: the \ac{BRP} and the \ac{BSP}. The former procures the resource's baseline power consumption; the latter sells the flexible resource's deviation from its baseline in flexibility markets. Only \acp{BRP} can trade power in intra-day markets. Therefore, \acp{BSP} must collaborate with \acp{BRP} to access such short-term markets, which proved to be complex in Continental Europe \cite{verbeij2023}. Alternatively, flexible resources can adapt their power consumption in cases where they do not keep power reserves, but this creates a rebound in power consumption. For instance, in \cite{TINA2022}, a commercial building's heating system participates in flexibility markets, with power rebound as the only form of adaptation.  

While both configurations appear separately in the literature, no research compares both scenarios, i.e., assesses the impact of trading on intra-day markets. Besides, the related literature neglects the thermal building model inaccuracy when quantifying a system's flexibility, at the risk of overestimating its potential.

\subsection{The Value of Thermal Comfort}

A building's heating system flexibility strongly depends on inhabitant preferences. Indeed, users commonly specify a range of indoor temperatures for which they feel comfortable. Field experiments, e.g.,  \cite{cowi2013} and \cite{Tomat2023}, identify that a 2\textdegree C indoor temperature range is deemed acceptable for inhabitants. A 3\textdegree C range could be accepted by inhabitants but yields discomfort and should be financially compensated \cite{cowi2013}. Field experiments also conclude that indoor thermal comfort is key to demand-side flexibility success. Indeed, if repeated discomfort is experienced by users,  consumers may disengage from flexibility programs \cite{sweetnam2019}. 

Flexibility quantification needs to account for thermal comfort. In the flexibility quantification, comfort constraints are imposed, usually in the form of soft constraints, in which violation is heavily penalized \cite{BUNNING2022, AMADEH2022}. Yet, constraint violations may still occur in operation \cite{GORECKI2017229} but are rarely accounted for in the economic valuation of flexibility. However, attributing a cost to the experienced discomfort that results from flexibility provision is necessary to account for consumers withdrawing from flexibility programs.

\subsection{Contributions}

Given the identified gaps in the literature review, we describe our contributions as follows: 

\begin{itemize}
    \item We develop an uncertainty-aware flexibility quantifier incorporating affine feedback policies to describe real-time baseline adaptation of a flexible resource. Since computing optimal feedback policies substantially increases the computation time, we additionally explore using sub-optimal fixed policies. 
    \item We examine the different flexibility quantification methods in two scenarios: a flexible resource can adapt its baseline power consumption in intra-day markets or rebound outside the flexibility period. 
    \item We suggest a discomfort cost procedure, aiming to attribute an economic cost to users' discomfort. We further demonstrate the necessity to account for such costs to compare formulations in a fair manner.
    \item We test our framework using measured data of an existing occupied residential building.   
\end{itemize}

\begin{figure}
    \centering
    \includegraphics[width = \columnwidth]{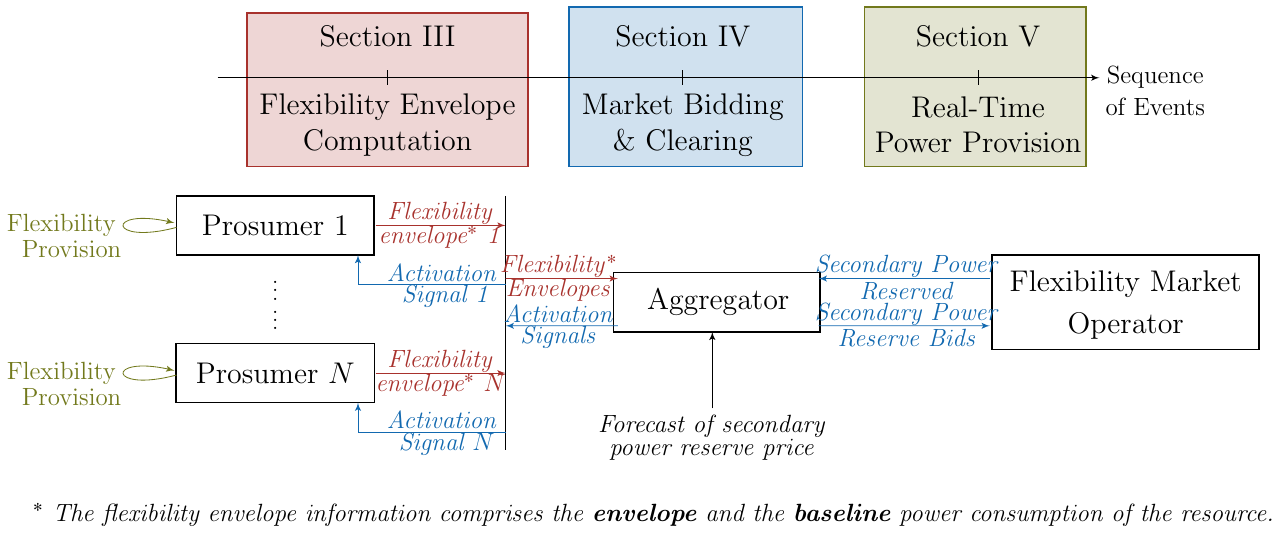}
    \caption{Flowchart describing the exchange of information between the stakeholders and the organization of the paper.}
    \label{fig:InformationFlowAFRR}
    \vspace{-0.3cm}
\end{figure}

 The rest of the paper is organized as follows. Section~\ref{sec:resBuilding&uncertainties} describes the modeling of the flexible residential building, emphasizing the uncertainties affecting room temperature predictions. Then, as Fig.~\ref{fig:InformationFlowAFRR} depicts, Section \ref{sec:flexEnvelope}, \ref{sec:bidding&clearing}, and \ref{sec:flexProvision} respectively describe the flexibility quantification in the form of energy flexibility envelopes, the participation in flexibility markets, and the flexibility provision in real-time. Section \ref{sec:caseStudy} describes the residential building and flexibility market chosen as a case study, and Section \ref{sec:results} discusses the results. Finally, Section \ref{sec:conclusion} concludes the work.  

\subsection{Notations}

In the remainder of this paper, bold letters designate vectors or matrices. The notation $\pmb{z}_t$ indicates the value of the variable $\pmb{z} \in \mathbb{R}^{N_z}$ at time instant $t$. The variable $\pmb{z}$ without time index denotes the matrix containing the collection of $\pmb{z}$ over a horizon of length $N$, i.e., $\pmb{z} = \left[ \pmb{z}_0, \ldots, \pmb{z}_N \right]^\intercal$. Overlines and tildes are used to denote nominal values and stochastic variations from such values, respectively.

\section{Building Modeling and Uncertainties}
\label{sec:resBuilding&uncertainties}

To describe the flexibility potential of a building's heating system, we must describe the building's thermal dynamics, i.e., the indoor temperature response to heating power inputs and ambient weather conditions. However, a building's thermal dynamics is uncertain, as it is subject to uncertain ambient weather forecasts and relies on a thermal model with limited accuracy. In this section, we describe the building thermal model and the associated ambient weather forecast and modeling uncertainties.

\subsection{Building Thermal Dynamics Model}

In this paper, we use a low-order linear model, which computes the room temperatures as a linear function of the past outdoor air temperatures, solar irradiances, and heating power inputs. We use a state-space representation of the building's thermal dynamics as follows: 
\begin{equation}
    \left\{
    \begin{aligned}
    \pmb{x}_{t+1} & = \mathbf{A} \pmb{x}_{t} + \mathbf{B} \pmb{u}_{t} + \pmb{\Tilde{w}}_{t},\\
    \pmb{y}_{t} & = \mathbf{C} \pmb{x}_{t} + \mathbf{D} \pmb{u}_{t} + \pmb{\Tilde{v}}_{t},
    \end{aligned}
    \right.
    \label{eq:subID}
\end{equation}
where $\left(\mathbf{A}, \mathbf{B}, \mathbf{C}, \mathbf{D} \right)$ are the system matrices, $\pmb{x}_t \in \mathbb{R}^{N_x}$ is the state vector and $\pmb{y}_t$ is the output vector, in our case, the room temperatures. Vector $\pmb{u}_t = \left[ \pmb{d}_t^\intercal,  \pmb{p}_t^\intercal \right]^\intercal$ contains the system's inputs, i.e., the uncontrollable weather variables $\pmb{d}_t$ and the controllable heating power inputs $\pmb{p}_t$. Hence, the input matrices $\mathbf{B} = \left[ \mathbf{B}_d, \mathbf{B}_p \right]$ and  $\mathbf{D} = \left[ \mathbf{D}_d, \mathbf{D}_p \right]$ can be decomposed into weather and heating inputs components.
Additionally, the linear model is subject to noise where $\pmb{\Tilde{w}}_{t} \sim \mathcal{N} \left(\pmb{0}, \pmb{\Sigma}_w \right)$ and $ \pmb{\Tilde{v}}_{t} \sim \mathcal{N} \left(\pmb{0}, \pmb{\Sigma}_v \right)$ are the process and measurement stochastic noises respectively and are assumed to be stationary-centered Gaussian. They are mutually independent and independent across timesteps.

We first derive, at time instant $t$, a prediction $k$-timesteps ahead of the system's behavior in response to ambient conditions and heating power inputs. For better readability, we introduce the matrices $\left( \pmb{\Lambda}^d_{k,i}, \pmb{\Lambda}^p _{k,i} \right)$ representing the impact of the past weather conditions and heating power inputs at time $t+i$ on the room temperatures at the predicted time $t+k$. Their exact definition is obtained by iterating (\ref{eq:subID}) \cite{618ecfce-dba1-3257-a376-142efb39f487}. The expected future room temperatures, $\Bar{\pmb{y}}_{t+k}$, follow:  
\begin{equation}
\begin{aligned}
     \Bar{\pmb{y}}_{t+k} = \pmb{c}_{t+k} + \sum_{i = 0}^{N} \left( \pmb{\Lambda}^p_{k,i} \pmb{p}_{t+i} + \pmb{\Lambda}^d_{k,i} \pmb{d}_{t+i} \right).
\end{aligned}
\label{eq:deterministic_SS}
\end{equation}
$\pmb{c}_{t+k} = \mathbf{C} \mathbf{A}^k \pmb{x}_t$ describes the impact of initial room temperatures, where $\pmb{x}_t$ is an estimate of the state at time $t$.

However, the behavior of future room temperatures is subject to model inaccuracy, which cannot be ignored over a long prediction horizon. Indeed, the thermal dynamics model error $k$-timesteps ahead follows: 
\begin{equation}
    \pmb{\Tilde{e}}_k = \pmb{\Tilde{v}}_k + \sum_{i = 1}^{k} \mathbf{C} \mathbf{A}^{i-1} \pmb{\Tilde{w}}_{k-i} = \pmb{\Tilde{v}}_k + \sum_{i = 0}^{N} \pmb{\Lambda}^w_{k,i} \pmb{\Tilde{w}}_{i},
    \label{eq:SS_error}
\end{equation}
where the matrix $\pmb{\Lambda}^w_{k,i}$ represents the impact of the process noise at time $i$ on temperatures at time $k$.
As the process and measurement noises are stationary and mutually independent, the $k$-timestep ahead model error is a centered Gaussian variable. 
Furthermore, even though thermal dynamics model errors of different future timesteps are dependent, a linear combination of these errors is a centered Gaussian variable.

\subsection{Weather Forecasts}

Future room temperatures are influenced by current and future ambient weather conditions. For the latter, a forecast is required. In this work, we obtain weather forecasts of geographical regions from a global forecasting agent. In order to obtain a refined forecast at the desired location, we further apply a Kalman filter, as detailed in \cite{Rousseau2023}.

Weather predictions are uncertain by nature. A first-order auto-regressive process provides a good approximation of the prediction error \cite{Rousseau2023}. Hence, the $k$-timesteps ahead weather forecast error follows:  
\vspace{-0.2cm}
\begin{equation}
    \pmb{d}_{t+k} = \pmb{\Bar{d}}_{t+k} + \pmb{\Tilde{d}}_k, \quad \text{and} \quad \pmb{\Tilde{d}}_k = \pmb{\varphi}^k \pmb{\Tilde{d}}_0 + \sum_{i=0}^{k-1} \pmb{\varphi}^i  \pmb{\Tilde{n}}_{k-i},
    \vspace{-0.2cm}
    \label{eq:WeatherForecastError}
\end{equation}
where $\pmb{\Tilde{d}}_0 \sim \mathcal{N} \left(\pmb{0}, \pmb{\Sigma}_{d,0} \right)$ and  $\pmb{\Tilde{n}}_k \sim \mathcal{N} \left(\pmb{0}, \pmb{\Sigma}_d \right)$ are centered, stationary, and mutually independent Gaussian noises.
The weather forecast error only depends on $k$, i.e., how many time steps into the future we predict weather conditions.
Even though weather forecast errors at different predicted timesteps are dependent, a linear combination of such forecast errors is a centered Gaussian stochastic variable.

\subsection{Complete Building Model}
\label{sec:overallModelError}

When predicting the room temperature $k$-timestep ahead, uncertainties stemming from the weather forecast and the model inaccuracy may yield inaccurate predictions. Consequently, given the previous derivations, the $k$-timestep room temperatures can be described as stochastic variables: 
\vspace{-0.2cm}
\begin{equation}
    \pmb{y}_{t+k} = \underbrace{\pmb{c}_{t+k} + \sum_{i = 0}^{N} \left( \pmb{\Lambda}^p_{k,i} \pmb{p}_{t+i} + \pmb{\Lambda}^d_{k,i} \pmb{\Bar{d}}_{t+i} \right)}_{\pmb{\Bar{y}}_{t+k}} + \underbrace{\sum_{i = 0}^{N} \pmb{\Lambda}^d_{k,i} \pmb{\Tilde{d}}_{i} + \pmb{\Tilde{e}}_k}_{\pmb{\Tilde{y}}_{k}}.
    \label{eq:finalSS}
\end{equation}
Since $\pmb{\Tilde{y}}_k$ consists of a linear combination of independent Gaussian noises, it is a centered Gaussian variable.

\section{Flexibility Quantification}
\label{sec:flexEnvelope}

Based on a building's thermal model, we can determine its flexibility potential. This work describes flexibility by energy envelopes, which are introduced in this section. Furthermore, acknowledging that a predicted system's response is associated with uncertainties, we present an uncertainty-aware energy envelope formulation that may, however, lead to over-conservative results. Hence, we further introduce real-time feedback when quantifying energy envelopes with the help of affine feedback policies.

\subsection{Uncertainty-Ignorant Envelope}

Energy envelopes describe the minimum and maximum heating energy consumption of a building which fulfill the technical constraints of the heating device and guarantee inhabitants' thermal comfort over a given prediction horizon. They delimit the feasible energy consumption region but are insufficient to fully represent the building's flexibility as power is also constrained. In this paper, we consider the heating's electric power rating as constant over time, leading to simple upper and lower bounds on the power.

The two bounds of an energy envelope can be derived from solving two optimizations, i.e., the maximization and minimization of energy consumption subject to the system's constraints, respectively. At time $t$, the upper energy bound is characterized by a thermal power vector $\pmb{p}_{\text{up},t}$ solution of:
\begin{subequations}
\begin{align}
     \max_{\pmb{p}, \pmb{\gamma}}  & \quad \sum_{k = 0}^{N}  \omega_k \left( \pmb{p}_{t+k}^\intercal \cdot \pmb{1}_{N_p}\right)  - \lambda \left( \pmb{\gamma}_{k}^{+} + \pmb{\gamma}_{k}^{-}\right)^\intercal \cdot \pmb{1}_{N_y} , \\
   \text{s.t.} \hspace{0.3cm} & \pmb{x}_{t} = \Hat{\pmb{x}}_{t}, \label{eq:UB_det_0}\\
     & \text{Eq. (\ref{eq:deterministic_SS})}, \hspace{4.2cm}\forall k \in \mathcal{H}^+, \label{eq:UB_det_1}\\
      & \pmb{p}_{\text{max}} \geq \pmb{p}_{t+k} \geq \pmb{p}_{\text{min}}, \quad \hspace{2.05cm} \forall k \in \mathcal{H}, \label{eq:UB_det_2}\\
    &  \mathbf{T}_{\text{max}} + \pmb{\gamma}_{k}^{+} \geq \pmb{\Bar{y}}_{t+k} \geq \mathbf{T}_{\text{min}} - \pmb{\gamma}_{k}^{-}, \hspace{0.45cm} \forall k \in \mathcal{H}^+, \label{eq:UB_det_3}\\
    & \pmb{\gamma}_k^+, \pmb{\gamma}_k^- \geq \mathbf{0}, \quad \hspace{3.15cm} \forall k \in \mathcal{H}^+, \label{eq:UB_det_4} \\
    & \pmb{p}_{t+N+1} = \pmb{0},\label{eq:UB_det_5}
\end{align}  
\label{eq:upperBoundDet}
\end{subequations}
where $\mathcal{H}^+ = [0, \cdots, N+1]$ and $\mathcal{H} = [0, \cdots, N]$. The estimated initial state, denoted by $\Hat{\pmb{x}}_{t}$, is recurrently updated using the latest room temperature measurements. Besides, terminal heating power inputs $ \pmb{p}_{t+N+1}$ mostly impact room temperatures beyond the optimization horizon and are, therefore, enforced to be zero. Additionally, a decreasing factor $\omega_k = e^{-k/N}$ weights the objective function's power components and favorizes early consumption, providing the widest possible feasible energy region.   

The system is subject to two types of constraints. First, the heating system ratings limit its power consumption. Second, inhabitants' thermal comfort restricts the range of acceptable room temperatures. We introduce the slack variables $\left( \pmb{\gamma}_k^+,  \pmb{\gamma}_k^- \right)$ which ensure the optimization feasibility. However, violating thermal comfort is strongly penalized by a large weight $\lambda$.

\subsection{Uncertainty-Aware Envelope}

Since future room temperatures are uncertain, so are the energy flexibility bounds. Hence, we propose an uncertainty-aware energy envelope formulation similar to~\cite{Rousseau2023}. Using a chance-constrained reformulation of the comfort constraints, we ensure that the upper and lower energy bounds fulfill thermal comfort, with a probability higher than $1-\epsilon_{\text{\tiny C}}$. 

As indicated in Section~\ref{sec:overallModelError}, future room temperatures can be modeled as Gaussian variables. Consequently, an explicit reformulation of the chance constraints exists. The interested reader may find more details on this reformulation in \cite{Rousseau2023}. In short, it consists of tightening the thermal comfort region by a safety margin $\pmb{s}_k^{\text{ua}} (\epsilon_{\text{\tiny C}})$, defined as:
\begin{equation}
    \pmb{s}_k^{\text{ua}} (\epsilon_{\text{\tiny C}}) = \sqrt{\text{Var} \left(\pmb{\Tilde{y}}_k \right)} \pmb{q}(1-\epsilon_{\text{\tiny C}}) = \left\| \pmb{\alpha}_k \pmb{\Sigma}^{1/2} \right\| \pmb{q}(1-\epsilon_{\text{\tiny C}}),
    \label{eq:safetyUA}
\end{equation}
where the matrix $\pmb{\Sigma} = \text{diag} \left(\pmb{\Sigma}_{d,0}, \pmb{\Sigma}_{d}, \pmb{\Sigma}_{v}, \pmb{\Sigma}_{w} \right)$ is composed of the variances introduced in Section~\ref{sec:resBuilding&uncertainties}, $\pmb{q}$ denotes the standard Gaussian's quantile function, and $\pmb{\alpha}_k$ can be obtained from (\ref{eq:finalSS}) by computing the standard deviation of $\Tilde{\pmb{y}}_k$. At time $t$, the uncertainty-aware upper energy bound results from a thermal power vector $\pmb{p}^{\text{ua}}_{\text{up},t}$ solution of:
\begin{subequations}
\begin{align}
   \max_{\pmb{p}, \pmb{\gamma}} & \quad \sum_{k = 0}^{N}  \omega_k \left( \pmb{p}_{t+k}^\intercal \cdot \pmb{1}_{N_p} \right)  - \lambda \left( \pmb{\gamma}_{k}^{+} + \pmb{\gamma}_{k}^{-}\right)^\intercal \cdot \pmb{1}_{N_y}, \\
    \hspace{-0.3cm} \text{s.t.} \hspace{0.2cm} & \text{Constraints (\ref{eq:UB_det_0}) - (\ref{eq:UB_det_2}), (\ref{eq:UB_det_4}) -  (\ref{eq:UB_det_5})}, 
    \label{eq:UB_CC_0}\\
    &  \pmb{\Bar{y}}_{t+k} \leq \mathbf{T}_{\text{max}} - \pmb{s}_k^{\text{ua}} (\epsilon_{\text{\tiny C}})  + \pmb{\gamma}_{k}^{+}, \quad \forall k \in \mathcal{H}^+,  \label{eq:UB_CC_1} \\
    & \pmb{\Bar{y}}_{t+k} \geq \mathbf{T}_ {\text{min}} + \pmb{s}_k^{\text{ua}} (\epsilon_{\text{\tiny C}}) - \pmb{\gamma}_{k}^{-}, \quad \forall k \in \mathcal{H}^+. \label{eq:UB_CC_2}
\end{align}  
\label{eq:upperBoundCC}
\end{subequations}
The lower energy bound can be computed accordingly.

\subsection{Uncertainty-Aware Envelope with Optimal Affine Feedback Policies}

\label{sec:AFP_opt}

% feedback
In real-time, a controller, e.g., a model-predictive controller, can decide upon the system's heating baseline power consumption based on the latest indoor temperature measurements, introducing real-time feedback. However, modeling such real-time feedback is a-priori intractable, which is why we consider linear feedback, yielding a tractable formulation \cite{BenTak2004}. 
Furthermore, adapting to past states, e.g., room temperatures, is equivalent to adapting to past disturbances, e.g., aiming to compensate for weather forecast and model errors \cite{GOULART2006}. Hence, we model the system's heating power consumption as a linear adaptation to past disturbances in the form of a matrix $\mathbf{M}$: 
\begin{equation}
    \pmb{\Tilde{p}}_{t+k} = \pmb{p}_{t+k} + \mathbf{M}_{k} \pmb{\Tilde{r}}_{t+k-1},
    \label{eq:AFP}
\end{equation}
with $\pmb{\Tilde{r}}_{t+k} = \left[ \pmb{\Tilde{d}}_{t+k}^\intercal; \pmb{\Tilde{e}}_{t+k}^\intercal \right]^\intercal$. The heating power consumption's adjustment at time $t+k$ depends on the latest measurement obtained at time $t+k-1$. Moreover, since $\pmb{\Tilde{r}}_{t+k-1}$ contains the accumulated disturbances up to timestep $t+k-1$, it is sufficient to adapt to its last observed instance.

We can now leverage this formulation in the computation of a chance-constrained uncertainty-aware energy bound including real-time adaptation. According to~(\ref{eq:AFP}), future heating power inputs are stochastic, resulting in two chance constraints. First, future room temperatures are stochastic, leading to thermal comfort chance constraints. Then, power constraints must be transformed into: 
\begin{equation}
    \mathbb{P} \left( \pmb{p}_{\text{max}} \geq \pmb{\Tilde{p}}_{t+k} \geq \pmb{p}_{\text{min}} \right) \geq 1-\epsilon_{\text{\tiny T}},
    \label{eq:chanceConstrainedPower}
\end{equation}
where $1-\epsilon_{\text{\tiny T}}$ describes the level of confidence imposed to technical constraints. Contrary to thermal constraints, technical constraints are hard constraints. Hence, we set a smaller value to the technical confidence level, $\epsilon_{\text{\tiny T}}$, than to the thermal comfort confidence level, $\epsilon_{\text{\tiny C}}$.

In this work, both chance constraints can be analytically reformulated. A safety factor $\pmb{s}_k^{\text{uaf},p} (\mathbf{M}_k, \epsilon_{\text{\tiny T}})$ reduces the feasible power region, ensuring that enough power is available to adapt to disturbances, with a high probability: 
\vspace{-0cm}
\begin{equation}
    \pmb{s}_k^{\text{uaf},p} (\mathbf{M}_k, \epsilon_{\text{\tiny T}}) = \underbrace{\left\| \mathbf{M}_k \pmb{\alpha}_{p,k} \pmb{\Sigma}^{1/2} \right\|}_{\sqrt{\text{Var} \left( \Tilde{\pmb{p}}_{t+k}\right)}} \pmb{q}\left(1-\epsilon_{\text{\tiny T}}\right),
    \vspace{-0.1cm}
\end{equation}
where $\pmb{\alpha}_{p,k}$ can be obtained by computing the standard deviation of $\Tilde{\pmb{p}}_{t+k}$. A second safety factor $\pmb{s}_k^{\text{uaf},c} (\mathbf{M}, \epsilon_{\text{\tiny C}})$ limits the feasible temperature range to maintain comfort in the presence of uncertainties: 
\vspace{-0cm}
\begin{equation}
    \pmb{s}_k^{\text{uaf},c} (\mathbf{M}, \epsilon_{\text{\tiny C}}) = \underbrace{\left\| \left( \pmb{\alpha}_k + \pmb{\alpha}_{1,k} \mathbf{M} \pmb{\alpha}_{2,k} \right) \pmb{\Sigma}^{1/2} \right\|}_{\sqrt{\text{Var} \left( \Tilde{\pmb{y}}_{t+k}\right)}} \pmb{q}(1-\epsilon_{\text{\tiny C}}),
    \vspace{-0.1cm}
\end{equation}
where the matrices $\pmb{\alpha}_{1,k}$ and $\pmb{\alpha}_{2,k}$ can be obtained by computing the standard deviation of $\Tilde{\pmb{y}}_{t+k}$ in the presence of affine feedback. A proper choice of linear feedback $\mathbf{M}$ can reduce the margin for the comfort constraints, compared to~(\ref{eq:safetyUA}), at the cost of reducing the feasible power region. In other words, we must trade off the reduced impact of accumulating uncertainties on room temperatures with the increased need for power margins to face uncertainties. 

Finally, the uncertainty-aware upper energy bound with feedback is associated with the power vector $\pmb{p}^{\text{uaf}}_{\text{up},t}$ and the optimal policy $\mathbf{M}_{\text{up},t}$, solutions of: 
\vspace{-0.2cm}
\begin{subequations}
\begin{align}
    & \hspace{-1cm} \max_{\pmb{p}, \pmb{\gamma}, \mathbf{M}}  \quad \sum_{k = 0}^{N}  \omega_k \left( \pmb{p}_{t+k}^\intercal \cdot \pmb{1}_{N_p}\right)  - \lambda \left( \pmb{\gamma}_{k}^{+} + \pmb{\gamma}_{k}^{-}\right)^\intercal \cdot \pmb{1}_{N_y} , \\
    \text{s.t.} \quad & \text{Constraints (\ref{eq:UB_det_0}) - (\ref{eq:UB_det_1}),(\ref{eq:UB_det_4}) - (\ref{eq:UB_det_5})}, \label{eq:UB_AFP_1} \\
    & \pmb{p}_{t+k} \leq \pmb{p}_{\text{max}} - \pmb{s}_k^{\text{uaf},p} (\mathbf{M}_k, \epsilon_{\text{\tiny T}}), \hspace{0.85cm} \forall k \in \mathcal{H},  \label{eq:UB_AFP_2} \\
    & \pmb{p}_{t+k} \geq \pmb{p}_ {\text{min}} + \pmb{s}_k^{\text{uaf},p} (\mathbf{M}_k, \epsilon_{\text{\tiny T}}) , \hspace{0.85cm} \forall k \in \mathcal{H}, \label{eq:UB_AFP_3} \\
    & \pmb{\Bar{y}}_{t+k} \leq \mathbf{T}_{\text{max}} - \pmb{s}_k^{\text{uaf},c} (\mathbf{M}, \epsilon_{\text{\tiny C}})  + \pmb{\gamma}_{k}^{+}, \hspace{0.0cm} \forall k \in \mathcal{H}^+,  \label{eq:UB_AFP_4} \\
    & \pmb{\Bar{y}}_{t+k} \geq \mathbf{T}_ {\text{min}} + \pmb{s}_k^{\text{uaf},c} (\mathbf{M}, \epsilon_{\text{\tiny C}}) - \pmb{\gamma}_{k}^{-}, \hspace{0.0cm} \forall k \in \mathcal{H}^+. \label{eq:UB_AFP_5}
\end{align}  
\label{eq:upperBoundAFP}
\end{subequations}
Both safety factors depend on the feedback matrix, i.e., an optimization variable. Hence, the uncertainty-aware linear optimization formulation (\ref{eq:upperBoundCC}) becomes a second-order cone convex optimization, increasing the computation time. 

The lower energy bound is computed similarly and is associated with the optimal feedback policy $\mathbf{M}_{\text{down},t}$.

\subsection{Uncertainty-Aware Envelope with Fixed Feedback}

Introducing affine feedback policies promises to reduce the conservativeness of energy bounds at the cost of increasing the computation time. Sub-optimal affine feedback policies offer an alternative solution to reduce computation time while modeling real-time feedback. Indeed, instead of computing the optimal policy at every iteration as described in Section~\ref{sec:AFP_opt}, we may employ an estimated affine feedback policy computed offline. With such a fixed sub-optimal linear feedback policy, the second-order cone optimization (\ref{eq:upperBoundAFP}) becomes the following linear optimization: 
\vspace{-0.1cm}
\begin{subequations}
\begin{align}
    & \hspace{-1cm} \max_{\pmb{p}, \pmb{\gamma}}  \quad \sum_{k = 0}^{N}  \omega_k \left( \pmb{p}_{t+k}^\intercal \cdot \pmb{1}_{N_p} \right) - \lambda \left( \pmb{\gamma}_{k}^{+} + \pmb{\gamma}_{k}^{-}\right)^\intercal \cdot \pmb{1}_{N_y}, \\
    \hspace{-0.1cm} \text{s.t.} \quad & \text{Constraints (\ref{eq:UB_AFP_1}) - (\ref{eq:UB_AFP_5})}, \label{eq:UB_AFPf_1} \\
    & \mathbf{M} = \mathbf{M}_{\text{up},t}^f,
    \vspace{-0.3cm}
\end{align}  
\label{eq:upperBoundAFPfixedPlicy}
\end{subequations}
where $\mathbf{M}_{\text{up},t}^f$ is the fixed affine upper bound policy computed offline. In the following, we discuss two different fixed policies for the upper and lower energy bounds. 

\subsubsection{Average Off-Line Policies}
\label{sec:avgPolicy}

Assuming that optimal affine feedback policies vary little with initial and ambient conditions, an average feedback policy approximates the optimal policy well \cite{OldewurtelStochastic2013}. In practice, we obtain such policies by solving~(\ref{eq:upperBoundAFP}) and its lower bound counterpart for a few training days contained in $\mathcal{N}_f$. In real-time, the average upper and lower bound feedback matrices serve as approximations of the optimal feedback policies:
\vspace{-0.2cm}
\begin{equation}
    \mathbf{M}_{\bullet,t}^{f, \text{avg}} = \frac{1}{|\mathcal{N}_f|} \sum_{i \in \mathcal{N}_f} \mathbf{M}_{\bullet,i, h_t}, 
    \vspace{-0.3cm}
\end{equation}
where $\bullet$ designates the indices $\text{\scriptsize up}$ and $\text{\scriptsize down}$. $\mathbf{M}_{\bullet,i,h_t}$ denotes the feedback matrix on day $i$, starting at hour $h_t$, where $h_t$ is the hour of time instant $t$. Hence, different average policies are used depending on the hour of $t$. 

\subsubsection{Cluster-Based Off-Line Policies}

Since a building is exposed to ambient weather, affine feedback policies may depend on outdoor conditions. Therefore, we suggest a cluster-based approach: first, days are clustered based on their weather conditions; then, the optimal feedback policy of the cluster center, computed offline, serves as the sub-optimal fixed policy.
This paper uses the standard K-means algorithm to form optimal clusters that minimize members' distance to their cluster center. Different clusters are used depending on the hour of the time instant $t$. 

\section{Flexibility Bidding in Reserve Markets}
\label{sec:bidding&clearing}

Energy flexibility envelopes describe the flexibility potential of resources that can be sold in a flexibility market. An aggregator receives the energy flexibility envelopes from a set of resources and aggregates their flexibility into one bid. The aggregation of resources into one envelope is out of the scope of this paper. This section only formulates the participation of such an aggregator in a flexibility market, given its aggregated envelope described by $\mathbf{E}_{\text{up}}$ and $\mathbf{E}_{\text{down}}$.

This paper assumes that the aggregator participates in a single reserve market. Such a market is sequential: first, reserves are purchased to ensure the availability of resources; then, in real-time, reserved resources receive a flexibility activation signal smaller than or equal to the reserved power. However, when reserves are procured, the activation signal is unknown, and the aggregator must guarantee resources' availability for all scenarios, including the worst case. Hence, we formulate a robust bidding procedure, considering the worst case, i.e., full activation of the reserved power. Besides, the aggregated flexibility is assumed to represent a small share of the flexibility market, and the aggregator can, thus, be considered a price-taker.

A resource's flexibility potential describes how it can adapt its consumption upon request. Therefore, it corresponds to the deviation of a resource's power consumption from its baseline consumption, denoted by the vector $\pmb{p}_b$. For simplicity, we assume that the aggregator knows the power consumption baselines of the resources in its portfolio.

Given some upward and downward reserve price forecasts, denoted as $\pmb{r}^{+}$ and $\pmb{r}^{-}$, respectively, the aggregator aims to maximize their revenues, formulated as:
\begin{subequations}
\vspace{-0.1cm}
\begin{align}
    &  \hspace{-0.3cm} \max_{\pmb{p}^{+}, \pmb{p}^{-}}  \quad \sum_{k = 1}^{N} r_{t+k}^{+} \left( \pmb{p}_{t+k}^{+\intercal} \cdot \pmb{1}_{N_p} \right) + r_{t+k}^{-} \left( \pmb{p}_{t+k}^{-\intercal} 
    \cdot \pmb{1}_{N_p} \right), \\
    & \hspace{-0cm} \text{s.t.} \nonumber \\
      & \pmb{p}_{\text{max},k} \geq \pmb{p}_{b,t+k} + \pmb{p}_{t+k}^+ \geq \pmb{p}_{\text{min},k}, \quad \hspace{1cm} \forall k \in \mathcal{H}, \label{eq:constraintPowerUpBidding}\\
     & \pmb{p}_{\text{max},k} \geq \pmb{p}_{b,t+k} - \pmb{p}_{t+k}^- \geq \pmb{p}_{\text{min},k}, \quad \hspace{1cm} \forall k \in \mathcal{H}, \label{eq:constraintPowerDownBidding}
     \\
     & \mathbf{E}_{\text{up},k} \geq \Delta t \sum_{i = 1}^{k} \left( \pmb{p}_{b,t+i} + \pmb{p}_{t+i}^+ \right) \geq \mathbf{E}_{\text{down},k},  \hspace{-0.05cm} \forall k \in \mathcal{H}, \label{eq:constraintEnergyUpBidding}\\
    & \mathbf{E}_{\text{up},k} \geq \Delta t \sum_{i = 1}^{k} \left( \pmb{p}_{b,t+i} - \pmb{p}_{t+i}^- \right) \geq \mathbf{E}_{\text{down},k}, \hspace{-0.05cm} \forall k \in \mathcal{H}, \label{eq:constraintEnergyDownBidding}
\end{align}  
\vspace{-0.2cm}
\end{subequations}

where $\pmb{p}^{+}$ and $\pmb{p}^{-}$ denote the upward and downward power reserves, respectively. These reserves must fulfill the robust power and energy constraints in both directions.

\section{Flexibility Provision and Adaptation}
\label{sec:flexProvision}

Based on the power reserved, resources receive a flexibility activation signal $\pmb{p}^s$ in real-time, requesting a change in power consumption. As discussed in Section~\ref{sec:introFeedback}, resources may adapt to realizing uncertainties in real time. For instance, resources can adapt their baseline power consumption in real-time, using intra-day markets \cite{BUNNING2022, GORECKI2017229}. We will refer to this situation as \textit{Scenario 1}. However, as outlined by \cite{TINA2022}, aggregators may be denied access to intra-day markets if they are not simultaneously \acp{BSP} and \acp{BRP}. In such cases, resources may only adapt their power consumption baseline when they do not provide reserves, also known as the rebound effect. We will refer to this second situation as \textit{Scenario 2}. The rest of this section describes the baseline adaptation of resources in both scenarios.

\subsection{Scenario 1: Intra-Day Market Participation}

Assuming access to intra-day markets, resources may adapt their baseline power consumption to new measured conditions. The power surplus or deficit compared to the initial baseline is then traded intra-day and is charged the intra-day market price. Specifically, a receding horizon controller is implemented at the resource level, determining the optimal baseline adaptation, given the new measured conditions and the future reserved flexible power: 
\vspace{-0.1cm}
\begin{subequations}
\begin{align}
    & \hspace{-1.1cm} \min_{\pmb{\Delta} \pmb{p}_b, \pmb{\gamma}} \hspace{0cm} \sum_{\scriptscriptstyle \bullet \in \{+,-\}}\sum_{k=1}^{N_r} r^{\bullet}_{\text{ID},t+k} \left( \pmb{\Delta} \pmb{p}^{\bullet \intercal}_{b,t+k} \cdot \pmb{1}_{N_p} \right) + \lambda \left(\pmb{\gamma}_{k}^{+} +  \pmb{\gamma}_{k}^{-}\right)^\intercal \cdot \pmb{1}_{N_y}, \nonumber\\ 
    % & \hspace{-0.3cm} \text{subject to:} \nonumber \\
     \text{s.t.} \hspace{0.2cm} & \text{Constraints~(\ref{eq:UB_det_0}), (\ref{eq:UB_det_4})}, \label{eq:realTimeController1_1}\\
    & \pmb{\Delta} \pmb{p}_{b, t+k} = \pmb{\Delta} \pmb{p}_{b, t+k}^+ - \pmb{\Delta} \pmb{p}_{b, t+k}^-, \hspace{0.9cm} \forall k \in \mathcal{H}, \\
    & \hspace{0cm} \pmb{p}^{\text{new},+}_{t+k} = \pmb{p}_{b,t+k}^0 + \pmb{\Delta} \pmb{p}_{b, t+k} + \pmb{p}^+_{t+k}, \hspace{0.4cm} \forall k \in \mathcal{H}, \label{eq:realTimeController1_2}\\
    & \text{Constraints~(\ref{eq:UB_det_1})-(\ref{eq:UB_det_3}) with } \pmb{p} = \pmb{p}^{\text{new},+}, \label{eq:realTimeController1_3}\\    
    & \hspace{0cm} \pmb{p}^{\text{new},-}_{t+k} = \pmb{p}_{b,t+k}^0 + \pmb{\Delta} \pmb{p}_{b,t+k} + \pmb{p}_{t+k}^-, \hspace{0.4cm} \forall k \in \mathcal{H}, \label{eq:realTimeController1_4}\\
    & \text{Constraints~(\ref{eq:UB_det_1})-(\ref{eq:UB_det_3}) with } \pmb{p} = \pmb{p}^{\text{new},-}, \label{eq:realTimeController1_5}\\
     & \pmb{\Delta} \pmb{p}^{+}_{b,t+k}, \pmb{\Delta} \pmb{p}^{-}_{b,t+k} \geq \pmb{0}, \hspace{2.2cm} \forall k \in \mathcal{H}. \label{eq:realTimeController1_6}
\end{align}  
\label{eq:realTimeControllerST}
\end{subequations}
The decision variable $\pmb{\Delta}\pmb{p}_b$ represents the optimal deviation from the initial baseline $\pmb{p}_{b}^{0}$ and is divided into a positive and negative part, charged with different short-term market prices $\pmb{r}^+_{\text{ID}}$ and $\pmb{r}^-_{\text{ID}}$, respectively. Both a reduction and an increase in the baseline incur a cost, so both intra-day market costs are positive. Besides, if the negative and positive power deviations are simultaneously non-zero, the system is charged a higher price than if at least one is zero for the same total power deviation value. Hence, there is no need to include a binary variable in this formulation. 

Formulation~(\ref{eq:realTimeControllerST}) is a robust receding horizon controller. It ensures, when adapting a resource's baseline, that the future promised power reserves are available with minimal discomfort even when all the reserved flexibility is activated. The optimization horizon $N_r$ designates the number of remaining timesteps until the end of the daily horizon and, hence, reduces as we move forward in time. For each iteration, only the first timestep's baseline deviation is traded on intra-day markets.  

\subsection{Scenario 2: Rebound Adaptation}

In actual power markets, aggregators may be denied access to intra-day markets. However, in current practices, aggregators can adapt their portfolio's baseline at timesteps when no reserve is kept. Such shift in the energy consumption is referred to as the rebound effect. We design a controller for individual resources to rebound based on recent conditions: 
\begin{subequations}
\begin{align}
    \hspace{-0.3cm} \min_{\pmb{\Delta} \pmb{p}_b, \pmb{\gamma}, \pmb{\beta}} & \hspace{0.2cm} \sum_{\scriptscriptstyle \bullet \in \{+,-\}}\sum_{k=1}^{N_r} r^{\bullet}_{\text{r}, t+k} \left( \pmb{\Delta} \pmb{p}^{\bullet \intercal}_{b, t+k} \cdot \pmb{1}_{N_p} \right) \nonumber \\
    & \hspace{0.6cm} + \alpha \left( \pmb{\beta}_k^\intercal \cdot \pmb{1}_{N_p} \right) + \lambda \left( \pmb{\gamma}^+_k +  \pmb{\gamma}^-_k \right)^\intercal \cdot \pmb{1}_{N_y} , \nonumber\\ 
    \text{s.t.} \hspace{0.2cm} & \text{Constraints~(\ref{eq:realTimeController1_1})-(\ref{eq:realTimeController1_6})}, \label{eq:realTimeControllerReb_1}\\
    & \pmb{\beta}_k = \pmb{0}, \hspace{2.9cm} \text{ if } \pmb{p}^{-}_{k} = \pmb{p}^{+}_{k} = \pmb{0}, \label{eq:realTimeControllerReb_2}\\
    & - \pmb{\beta}_k \leq \pmb{\Delta} \pmb{p}^{+}_{b,t+k} - \pmb{\Delta} \pmb{p}^{-}_{b,t+k} \leq \pmb{\beta}_k, \hspace{0.1cm} \text{otherwise}. \label{eq:realTimeControllerReb_3}
\end{align}  
\label{eq:realTimeControllerReb}
\end{subequations}
In timesteps without reserves, a positive or negative baseline deviation is charged with the price $r_{\text{r}}^+$ and $r_{\text{r}}^-$, respectively. However, in timesteps when reserves are scheduled, a deviation from the initial baseline is considered a default of flexibility provision and is strongly penalized with the weight $\alpha$. A strict rebound policy that forbids any power deviation when reserves are scheduled can be implemented by replacing~(\ref{eq:realTimeControllerReb_3}) with an equality constraint setting $\pmb{\Delta} \pmb{p}_b$ to zero. In this formulation, two strong penalizations compete: one discourages thermal discomfort while another penalizes the non-provision of the promised flexibility.

\section{Case Study}
\label{sec:caseStudy}

To assess the performances of the proposed methodology, we study the participation of an aggregator with a single flexible resource, namely the UMAR building, in the German secondary frequency control market. The methodology could be extended to aggregators with multiple assets, but is limited to a single resource in this paper for clarity.

\subsection{\ac{UMAR}}

\ac{UMAR} is a residential apartment integrated into the NEST experimental building at Empa campus \cite{Richner17}. It comprises two bedrooms and one large living room on a total of 155~m$^2$. \ac{UMAR} has large windows and is equipped with ceiling radiant heating panels in which hot water flows from a central heat pump. The hot water flow entering each room can be controlled with valves. \ac{UMAR}'s total thermal power capacity is 5~kW$_{\text{th}}$.

Additionally, the authors in \cite{Khayatian2022} developed a high-fidelity digital twin of \ac{UMAR}, named \texttt{nestli} and openly available on GitHub\footnote{\url{https://github.com/hues-platform/nestli}}. Based on the building software EnergyPlus, \texttt{nestli} simulates \ac{UMAR}'s indoor conditions in response to various inputs such as heating power inputs, window openings, and inhabitants' heat gains. \ac{UMAR}'s digital twin offers two key benefits in this work. First, it generates reliable historical data to identify the low-order building model and its associated uncertainty. Historical indoor conditions are replicated using measured historical heating power consumption, weather conditions, and internal gains, but with windows closed. Second, the controllers presented in Section~\ref{sec:flexProvision} can be implemented in \ac{UMAR}'s digital twin to assess their performance and the resulting thermal discomfort.

\subsection{The German Secondary Frequency Control Market}

Secondary frequency control, referred to as \ac{aFRR} in Continental Europe, designates an automatic control mechanism that returns the electric power grid frequency to its nominal value and restores tie-line power flows in interconnected systems after an active power imbalance. To ensure sufficient reserves in operation, \acp{TSO} first procure reserves in a reserve market and request a share of the reserved power in operation, according to the measured power system imbalance.

We assume that the heating system of the residential building \ac{UMAR} participates in the German \ac{aFRR} market. Even though UMAR is located in Switzerland, the German \ac{aFRR} market is preferred over the Swiss market as German reserves are purchased every 4~hours \cite{BalncingRulesGermany} compared to a weekly procurement in Switzerland. This granular procurement yields peak upward\footnote{The upward direction designates an increase in production, or decrease in consumption, while the downward direction refers to a decrease in production or an increase in consumption.} prices at peak hours to decrease consumption and peak downward prices at off-peak hours to increase consumption. Furthermore, the German \ac{aFRR} energy and reserve revenues display comparable values.

To ensure the system's reliability in operation, TSOs must purchase reserves to support large component outages and ensure the $N-1$ security requirements. Yet, in operation, such outages are rare, leading TSOs to only request no or a small share of the reserved power most of the time. Consequently, the utilization rate  defined as the average ratio of the requested to the reserved power is small. In Germany, it equals about 3\%. 

In this paper, we assume that the German \ac{aFRR} market is cleared daily at midnight. The delivery starts immediately after the market clearing. We further assume that the reserve market has an hourly resolution. 

If the reserved power is requested but not provided by flexible resources, a strong penalty cost applies. It aims to prevent flexible resources from malpractice and amounts to the imbalance price \cite{BalncingRulesGermany}. A slack of 5\% around the requested value is allowed by grid operators.

\subsection{Continuous Intra-Day Markets}

Continental Europe's continuous intra-day markets are short-term markets, where power quantities are traded shortly before the physical power delivery. There is no central clearing in such markets, but individual matching offers are cleared continuously. Before the physical power delivery, market participants can trade intra-day based on updated predictions of their production or consumption. 

In \textit{Scenario 1}, resources may adapt their baseline power consumption in intra-day markets. This paper assumes that all bids placed by the aggregator are cleared. Furthermore, intra-day prices are likely close to day-ahead prices \cite{andrade2017}. This paper assumes that a fee of 20\% of the day-ahead price is incurred for intra-day trades. This may represent additional transaction fees \cite{MERTEN2020114951}, but also a pessimistic deviation from the day-ahead price. Hence, every intra-day trade is charged at an additional 20\% of the day-ahead price.

In \textit{Scenario 2}, resources may only adapt their baseline when no power is reserved, causing a rebound effect. This paper assumes that the rebound energy is charged at an additional 20\% of the day-ahead price, similar to intra-day trades, allowing for a fair comparison between formulations.

\section{Results and Discussion}
\label{sec:results}
Using the UMAR building as an example, we assess the performance of the approaches developed in this paper.
We first validate the proposed uncertainty modeling by comparing it to observed errors. Then, we compare different envelope formulations and highlight how affine feedback policies can mitigate the impact of future uncertainties. Finally, we analyze the participation of the flexible UMAR building in an \ac{aFRR} market.

\subsection{Uncertainty Modelling}

Uncertainties are at the core of the uncertainty-aware energy flexibility quantification. This paper considers two sources of uncertainties: weather forecasts and building thermal modeling inaccuracies. Based on historical data, both error distributions are analyzed for various prediction horizons, ranging from 1 to 24 timesteps (one day-ahead). As the uncertainty modelling of weather forecast errors has already been discussed in \cite{Rousseau2023}, we focus on uncertainties associated with the thermal modelling inaccuracy.

\begin{figure}
    \centering
    \includegraphics[width = \columnwidth]{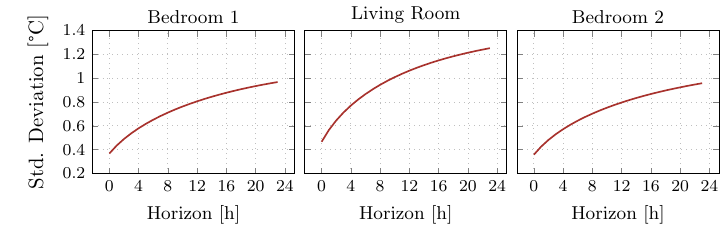}
    \caption{Comparison between the measured model error and the state-space representation of the model error.}
    \label{fig:sigmaComparisonStateSpace}
     \vspace{-0.3cm}
\end{figure}

Fig.~\ref{fig:sigmaComparisonStateSpace} displays the variance of the error when predicting UMAR's room temperatures. One-timestep ahead, the variance of the error between the predicted and measured room temperatures is low and comparable to values reported in \cite{AMADEH2022}. However, when predicting room temperatures over longer horizons, the variance significantly increases. Based on the available data and the error model in~(\ref{eq:SS_error}), the error variance of the three rooms stabilizes to a long-term value.

Fig.~\ref{fig:sigmaComparisonStateSpace} also highlights the differences in modeling accuracy among the rooms. Indeed, the bedrooms' modeling is more accurate. As the living room also comprises a kitchen, more significant heat gains impact the room's thermal dynamics. Besides, because of large windows, solar irradiance strongly impacts the living room's dynamics, which the low-order model does not fully capture.  

\subsection{Flexibility Envelope Quantification}

To compare the different flexibility quantification formulations, we compute \ac{UMAR}'s energy flexibility potential over 24~hours for 20 random days, referred to as samples. The set of these samples is denoted as $\mathcal{S}_{\text{env}}^{20}$. These samples are randomly selected between Dec.~2020 and Feb.~2021, and the sample set is fixed across formulations.

\subsubsection{Envelope Formulations}

\begin{figure}
    \centering
    \includegraphics[width = \columnwidth]{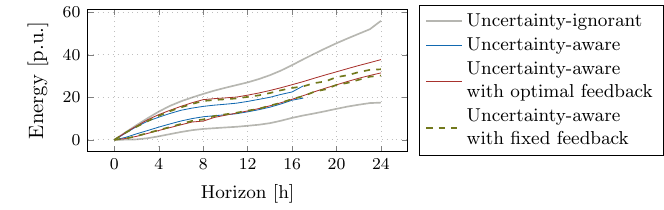}
    \caption{Average flexibility envelope of UMAR in the set $\mathcal{S}_{\text{env}}^{20}$ for the different formulations over a horizon of a day, with $1-\epsilon_{\text{\tiny C}} = 80\%$ and $\mathbf{T}_{\text{max}} - \mathbf{T}_{\text{min}} = 2$\textdegree C.}
    \label{fig:avgEnvelope80}
     \vspace{-0.3cm}
\end{figure}

The uncertainty-ignorant flexibility envelope determines the minimum and maximum energy that can be consumed by UMAR's heating system, neglecting uncertainties. Uncertainty-aware envelopes account for uncertain future ambient conditions and the inaccurate thermal model, reducing the flexibility potential. Fig.~\ref{fig:avgEnvelope80} compares the average envelope of  $\mathcal{S}_{\text{env}}^{20}$ for a comfort range of 2\textdegree C and a thermal comfort confidence of 80\%. The uncertainty-ignorant formulation largely overestimates UMAR's flexibility potential. As the flexibility prediction horizon increases, ambient conditions and the thermal model become more uncertain. Hence, the difference between the uncertainty-ignorant and uncertainty-aware flexibility potential increases.

Fig.~\ref{fig:avgEnvelope80} also highlights a limit in quantifying flexibility. In some cases, a \ac{MFPH} exists, which describes a maximum duration after which no flexibility can be provided with enough confidence. For instance, the \ac{MFPH} of the uncertainty-aware formulation is reached after 16~hours. After that point, the reduced maximum temperature limit becomes smaller than the increased minimum in the thermal chance constraint. Hence, flexibility cannot be guaranteed with sufficient confidence. 

Introducing feedback offers a less conservative alternative. As the residential building adapts its baseline power consumption, the effect of past uncertainties reduces. In other words, an affine policy reduces the safety margin applied to the comfort region. Fig.~\ref{fig:avgEnvelope80} displays a wider flexibility envelope with feedback compared to the uncertainty-aware approach without feedback. Additionally, introducing feedback can also extend the \ac{MFPH}, as Fig.~\ref{fig:avgEnvelope80} illustrates. This is particularly important if high flexibility revenues occur towards the end of the quantification horizon.

\subsubsection{Fixed Affine Policies}

Employing feedback policies may reduce the conservatism of the uncertainty-aware formulation. Yet, it also greatly increases the envelope computation time. Hence, we also consider using sub-optimal affine feedback policies computed offline. Only a limited number of affine feedback policies must be computed and can be directly integrated into the online optimization. 

Cluster-based policies assume a strong dependency of the feedback policies to the ambient weather conditions. Based on historical data from Dec.~2019 to Feb.~2020, days with similar ambient conditions are clustered. For each cluster, the affine feedback policy for the cluster center is then computed. An analysis of the evolution of the average and maximum entry-wise $\ell_2$ distance between the optimal and the fixed affine feedback policy indicates that distances vary little when increasing the number of clusters. This observation aligns with the fact that dominant uncertainties stem from the thermal modeling inaccuracy \cite{Rousseau2023}.

\begin{figure}
    \centering
    \includegraphics[width = 0.9\columnwidth]{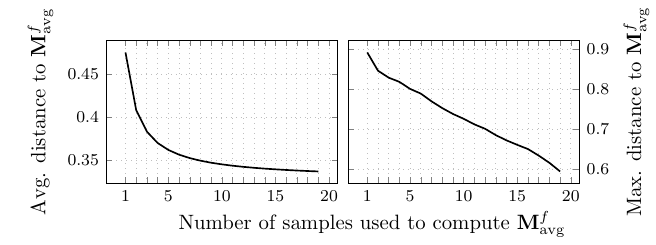}
    \caption{Average and maximum distance in $\mathcal{S}_{\text{env}}^{20}$ between the optimal affine policy $\mathbf{M}$ and the average one $\mathbf{M}^f_{\text{\scriptsize avg}}$, as a function of the number of samples used to compute $\mathbf{M}^f_{\text{\scriptsize avg}}$.}
    \label{fig:impactNbSamplesAVg}
     \vspace{-0.3cm}
\end{figure}

Since weather conditions only partially impact feedback policies, an average affine policy is also a promising sub-optimal choice. Fig.~\ref{fig:impactNbSamplesAVg} analyzes the performances of an average sub-optimal policy,
in terms of entry-wise $\ell_2$ distance between the individual optimal policies of  $\mathcal{S}_{\text{env}}^{20}$ and the average fixed policy, and for an increasing number of training samples used to construct the average policy, from 1 to 20. The distance to the optimal policies stabilizes, on average, after 10~training samples. Hence, adding more training samples to compute the average feedback policy does not significantly change the sub-optimal average matrix, on average. Nevertheless, the maximum distance continuously decreases as the number of training samples increases, indicating that more samples yield better performances.

Overall, the analysis reveals that an average fixed feedback policy yields better performance than a cluster-based approximation. Moreover, as Fig.~\ref{fig:impactNbSamplesAVg} indicates an improved performance with an increasing number of samples, we employ an average fixed feedback policy computed based on 20~training samples in the rest of this paper. 

\subsubsection{Advantages of Incorporating Feedback}
\label{sec:AdvantagesFeedback}

Fig.~\ref{fig:avgEnvelope80} illustrates the impact of employing a fixed average feedback policy on the flexibility envelope. While using fixed feedback (dashed line) slightly reduces the average envelope width compared to optimal feedback (red line), it still significantly increases the flexibility compared to the uncertainty-aware without feedback case. It also successfully extends the \ac{MFPH}, as does the formulation with optimal feedback. 

\begin{figure}
     \centering
     \begin{subfigure}[b]{0.8\columnwidth}
         \centering
        \includegraphics[width = \columnwidth]{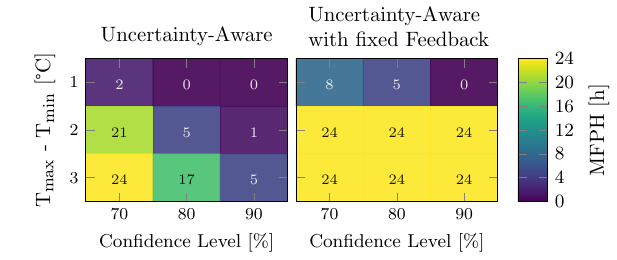}
        \caption{Maximum Flexibility Provision Horizon (MFPH).}
        \vspace{0.3cm}
        \label{fig:mfph}
     \end{subfigure}
     \hfill
     \begin{subfigure}[b]{0.8\columnwidth}
     \vspace{-0.3cm}
         \centering
        \includegraphics[width = \columnwidth]{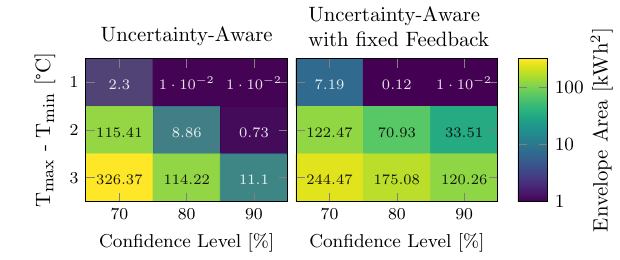}
        \caption{Flexibility Envelope Area (FEA).}
        \label{fig:envelopeArea}
     \end{subfigure}
        \caption{Comparison between the uncertainty-aware and the uncertainty-aware with fixed feedback formulations, in $\mathcal{S}_{\text{env}}^{20}$.}
        \label{fig:comparisonAware&afp}
         \vspace{-0.3cm}
\end{figure}

Fig.~\ref{fig:comparisonAware&afp} provides a more thorough comparison of two formulations: the uncertainty-aware without feedback and with fixed averaged feedback. Two metrics are used: the \ac{FEA}, describing the area between the two bounds, and the \ac{MFPH}. Implementing feedback effectively extends the \ac{MFPH} and increases the \ac{FEA} for most thermal comfort ranges and confidence levels.

However, Fig.~\ref{fig:comparisonAware&afp} also illustrates limits to the level of improvement that can be achieved by fixed feedback. Indeed, with a 3\textdegree C comfort range and a confidence level of 70\%, the flexibility potential is slightly higher in the uncertainty-aware case. The fixed affine policy creates stricter power constraints, which limits the flexibility potential. In this case, an optimal feedback policy approximation or more advanced fixed policy approximation should be sought.

\subsection{Flexibility Bidding and Provision}
Based on \ac{UMAR}'s flexibility potential, the aggregator bids into the German \ac{aFRR} reserve market. At midnight, the market is cleared for the coming day. Based on the reserved power, \ac{UMAR} receives flexibility requests. We assess the performances on a set of 6~random days, denoted as $\mathcal{S}_{\text{prov}}^{6}$, between Dec.~2020 and Feb.~2021. This limited number of days allows us to test various configurations while keeping the computation time low.  

\subsubsection{Metrics}

To compare the results, we introduce two metrics. First, we analyze the net revenues obtained by UMAR when participating in the German \ac{aFRR} market. Four types of revenues must be considered. \ac{UMAR} receives revenues to keep power reserves and is remunerated for the provided flexible energy. However, when adapting its baseline power consumption, it incurs intra-day or rebound costs. Besides, in extreme cases, if the occupants in \ac{UMAR} favor thermal comfort over flexibility provision, a penalty cost is charged. Summing up costs and revenues yields \ac{UMAR}'s net flexibility revenue. 

However, economic revenues must always be considered together with thermal discomfort, as a higher revenue may also incur a larger discomfort for inhabitants. Hence, we also evaluate the incurred temperature violations. Revenues are given as per-unit values, normalized by the largest average revenue obtained in aFRR markets with access to intra-day markets, and thermal discomfort as temperature deviations from the upper or lower bound.

\subsubsection{Scenario 1 - Intra-day Market Access}

\begin{figure}
     \centering
     \includegraphics[width = \columnwidth]{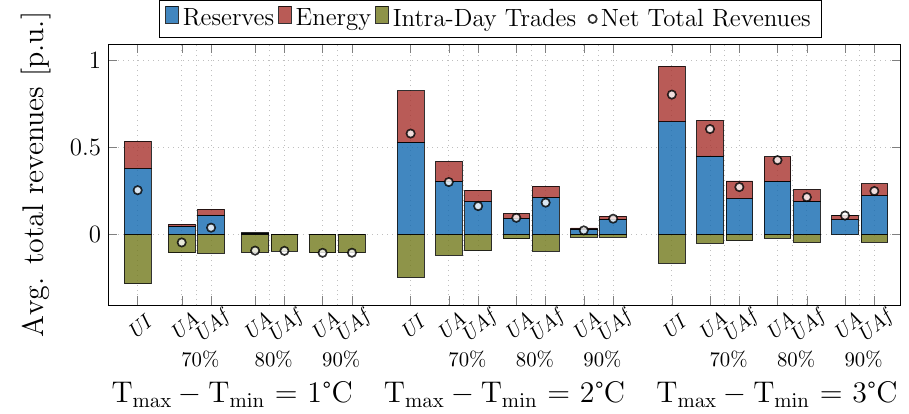}
    \caption{Average daily revenues of UMAR in \ac{aFRR} markets, over $\mathcal{S}_{\text{prov}}^{6}$, with intra-day market acces (\textit{Scenario 1})\protect\footnotemark.}
    \label{fig:barChart_STMarket}
     \vspace{-0.3cm}
\end{figure}
\begin{figure}
    \centering
    \includegraphics[width = 0.9\columnwidth]{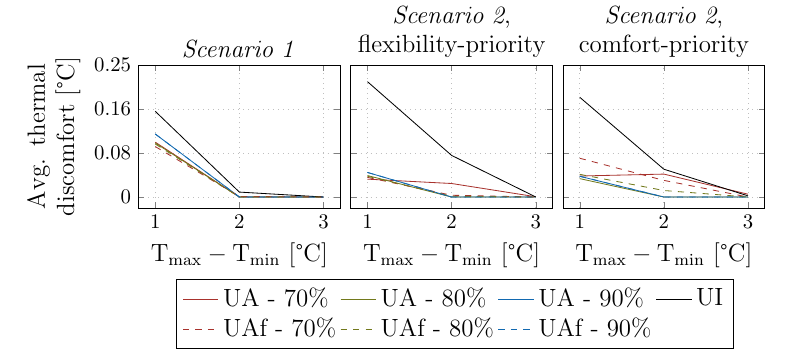}
    \caption{Average thermal discomfort in UMAR over $\mathcal{S}_{\text{prov}}^{6}$ during secondary control provision$^3$.}
    \label{fig:thermalViolationVDdeltaT}
\end{figure}

\footnotetext{UI stands for Uncertainty-Ignorant, UA for Uncertainty-Aware, and UAf for Uncertainty-Aware with Fixed Feedback. This remark holds for Fig.~\ref{fig:barChart_STMarket} to~\ref{fig:revenueVSratio}.} 

Fig.~\ref{fig:barChart_STMarket} illustrates UMAR's average revenues if continuous intra-day trades are allowed. The uncertainty-ignorant envelope overestimates \ac{UMAR}'s flexibility potential. Consequently, it yields large reserve and energy revenues. In return, more energy is traded intra-day. However, intra-day costs remain low enough to yield positive net revenues (white dot) with an uncertainty-ignorant flexibility quantification method. 

An uncertainty-aware flexibility quantification significantly reduces the flexibility potential. Therefore, it reduces the reserve and energy revenues. In return, smaller amounts of energy are traded intra-day. In some cases, introducing fixed feedback policies in the quantification substantially increases the reserve and energy revenues while keeping low intra-day costs. For instance, for a thermal comfort range of 1\textdegree C and confidence of 70\%, including feedback in the flexibility quantification yields higher reserve and energy revenues while keeping the same intra-day costs. In this case, introducing feedback transforms negative net revenues into positive ones. In other cases, e.g., for a comfort range of 2\textdegree C and confidence of 80\%, introducing feedback policies yields lower revenues. The uncertainty-aware quantifier with feedback extends the \ac{MFPH} at the cost of reducing the flexibility amount. Yet, as large prices are not all allocated toward the end of the horizon, the late flexibility prices do not compensate for the loss in flexibility, leading to a decrease in revenues.

Fig.~\ref{fig:barChart_STMarket} also shows intra-day trades, even when no flexibility is provided, for a tight comfort range of 1\textdegree C. For instance, with a confidence level of 90\%, no flexibility can be provided with enough confidence. Hence, \ac{UMAR} earns no flexibility revenues. Yet, some energy is traded intra-day to maintain the room temperatures within the comfort range in the baseline case. Hence, high comfort requirements with narrow temperature ranges may be difficult to satisfy in the context of flexibility provision. 

Fig.~\ref{fig:thermalViolationVDdeltaT} displays the thermal discomfort associated with flexibility provision. It displays the average deviation from the thermal comfort bound. In \textit{Scenario 1}, the worst thermal discomfort occurs in the case of an uncertainty-ignorant flexibility quantification. Yet, average thermal discomfort levels remain comparable among the different formulations, especially for temperature ranges larger than 2\textdegree C. However, further analysis reveals that, for a thermal range of 1\textdegree C, the worst temperature discomfort over the 24~hour horizon is 0.4\textdegree C with an uncertainty-ignorant formulation, whereas it is smaller than 0.25\textdegree C for all other formulations. For a comfort range of 3\textdegree C, the thermal comfort of inhabitants is satisfied for all the flexibility quantification formulations.

\subsubsection{Scenario 2 - Rebound Power Adaptation}

\begin{figure}
     \centering
     \includegraphics[width = \columnwidth]{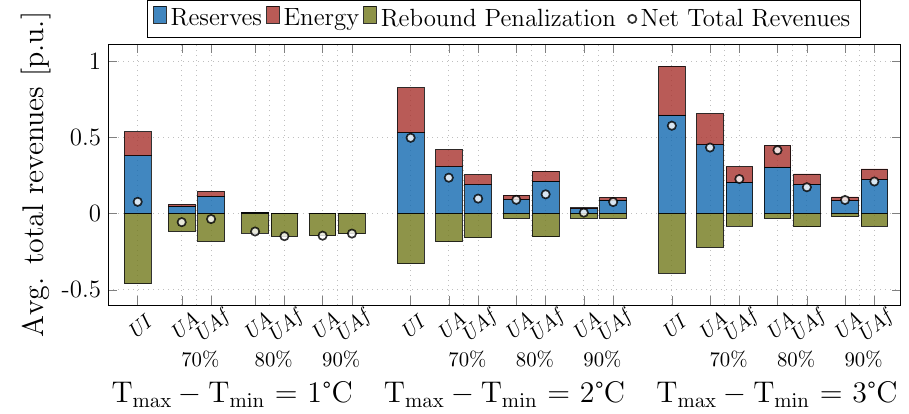}
    \caption{Average daily revenues of UMAR in \ac{aFRR} markets in \textit{Scenario 2}, with flexibility provision as the priority$^3$.}
    \label{fig:barChart_lowDiscomfort}
    \vspace{-0.3cm}
\end{figure}

For the case without intra-day markets, Fig.~\ref{fig:barChart_lowDiscomfort} displays UMAR's flexibility average net revenues over $\mathcal{S}_{\text{prov}}^{6}$. In this scenario, UMAR's baseline power consumption can still be adapted as power rebound, i.e., a change of the power baseline only when no flexibility is reserved. As the baseline can only be adapted at certain timesteps, it is less efficient than in \textit{Scenario 1}. The flexibility provision requires more energy to adapt the baseline, resulting in a rebound cost. As a consequence, net revenues are smaller in this scenario. It also benefits uncertainty-aware flexibility quantification, with and without feedback, which tends to use less energy to adapt baselines. Fig.~\ref{fig:barChart_lowDiscomfort} also highlights that flexibility provision tends not to be profitable for a 1\textdegree C comfort range for the considered building. 

Fig.~\ref{fig:thermalViolationVDdeltaT} displays the thermal discomfort of \ac{UMAR}'s inhabitants when providing flexibility in \textit{Scenario 2}. Since the baseline adaptation is less efficient, it also causes more violations of the thermal comfort bounds. In particular, an uncertainty-ignorant flexibility quantification leads to a significant increase in thermal discomfort when providing flexibility, compared to uncertainty-aware formulations.

Both the net revenues and the experienced thermal discomfort reflect the performances of different flexibility quantification formulations. Yet, in literature, methods tend to compare economic flexibility revenues, neglecting discomfort \cite{li2023unlocking}. To fairly compare methods, thermal discomfort should be reflected in the flexibility economic potential. Hence, we assess \textit{Scenario 2} when comfort is prioritized, i.e., if the thermal weighting factor $\lambda$ dominates over the flexibility provision weighting factor $\alpha$ in~(\ref{eq:realTimeControllerReb}). In this scenario, the baseline is adapted even when no power flexibility is reserved. When power flexibility is reserved, the building controller deviates from the requested power if such power consumption creates thermal discomfort, but this deviation comes at a high cost. 

\begin{figure}
     \centering
     \includegraphics[width = \columnwidth]{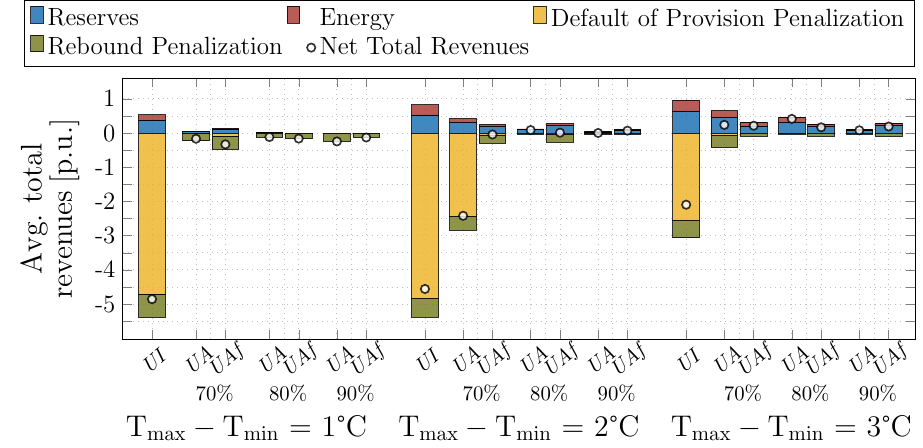}
    \caption{Average daily revenues of UMAR in \ac{aFRR} markets in \textit{Scenario 2}, with thermal comfort priority$^3$.}
    \label{fig:barChart_highDiscomfort}
     \vspace{-0.4cm}
\end{figure}

Fig.~\ref{fig:barChart_highDiscomfort} represents the net revenues obtained in \textit{Scenario 2}, when thermal comfort is deemed more important than power flexibility provision. An uncertainty-ignorant flexibility quantification overestimates the flexibility potential, which leads to thermal discomfort when delivering flexibility. With thermal comfort priority, this leads to deviations from the requested power in real-time and, thus, a large penalty. In comparison, uncertainty-aware formulations, with and without feedback, provide the requested power in real-time without violating thermal comfort. These formulations yield positive net revenues in most cases. For instance, for a comfort range of 2\textdegree C, an 80\%-uncertainty-aware flexibility quantification with feedback results in the highest total net revenue.

\subsubsection{Price Sensitivity}
\label{sec:discussion}

\begin{figure}
    \hspace{0.5cm}
    \includegraphics[width = 0.8\columnwidth]{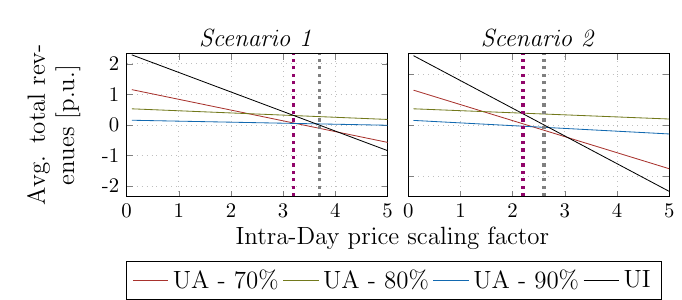}
    \caption{UMAR's average net flexibility revenues as a function of the intra-day price multiplicative factor, with $\mathbf{T}_{\text{max}} - \mathbf{T}_{\text{min}} = 2$\textdegree C (the uncertainty-aware formulations with feedback are omitted for readiness)$^3$.}
    \label{fig:revenueVSratio}
     \vspace{-0.3cm}
\end{figure}

Fig.~\ref{fig:revenueVSratio} shows the change of \ac{UMAR}'s flexibility revenues in \textit{Scenario 1} and \textit{2 with flexibility provision priority} when multiplying the intra-day prices by a multiplicative scaling factor, all other prices remaining unchanged. If intra-day prices are low, the uncertainty-ignorant quantification yields large flexibility revenues in both scenarios. However, as intra-day prices increase, adapting the baseline becomes expensive. The purple dashed line indicates the multiplicative factor after which another flexibility quantification formulation outperforms the uncertainty-ignorant one in terms of flexibility revenues, and the grey dashed line defines the profitability threshold of the uncertainty-ignorant formulation. For a comfort range of 2\textdegree C, intra-day prices must triple for an uncertainty-aware formulation to outperform the uncertainty-ignorant one in \textit{Scenario 1}, but only double in \textit{Scenario 2}. In \textit{Scenario 2}, both thresholds are lower than in \textit{Scenario 1}. As the baseline adaptation is less efficient in \textit{Scenario 2}, higher prices would reduce revenues faster. 

An additional analysis reveals that when the reserve price is low, the uncertainty-ignorant formulation is profitable only for low intra-day prices, while uncertainty-aware formulations yield positive net revenues already at low intra-day prices. If reserve prices are halved, uncertainty-aware formulations economically outperform the uncertainty-ignorant one already with the current energy and intra-day prices in \textit{Scenario 2}.

\section{Conclusion}
\label{sec:conclusion}

This paper explores the concept of uncertainty-aware energy envelopes to quantify the flexibility of buildings' heating systems, using a chance-constrained reformulation of thermal comfort constraints to account for weather forecasts and inaccuracies in the thermal model of a building. We further explore and provide methods for the consideration of real-time feedback in the quantification of the flexibility which reduces the long-term impact of uncertainties. Based on the case study of an existing building, we reveal that accounting for modeling inaccuracy is key to obtaining a reliable flexibility quantification. Besides, even though an uncertainty-aware potential offers a reliable estimate, it may also be over-conservative. We show that we can increase flexibility by modeling feedback in the quantification, assuming such feedback exists in operation. 

To exemplify the framework above, this paper additionally explores flexibility provision in flexibility reserve markets. The results reveal that, when intra-day trades are available, an uncertainty-ignorant quantification is preferable when participating in the German \ac{aFRR} market characterized by an infrequent activation of the reserved power. However, if resources can only adapt their power consumption when no flexibility is reserved, an uncertainty-aware formulation may be preferred to avoid repeated thermal discomfort. Economically speaking, such a formulation is preferred by comfort-prioritizing inhabitants.

The results presented in this paper highlight the need to account for thermal comfort in the economic evaluation of different flexibility quantification formulations. Indeed, while comfort is respected in the quantification, it may be violated in real-time operation, as shown in this paper. Hence, assigning a cost to thermal discomfort is needed for a fair comparison of the methods. While we provide a case study based on current strict German penalization rules, we believe such a cost, even with a lower value, cannot be neglected when evaluating the results. 

Future works should apply the methodology presented in this paper to different buildings to evaluate the transferability of the proposed method, especially to buildings differently affected by uncertainties, e.g., with less stochastic heat gains from users. 
% Besides, future large-scale flexibility quantification also necessitates a clarification of the role of an aggregator, specifically its aggregation and disaggregation strategy for heterogeneous resources. 
Besides, this paper only exemplifies flexibility use in the context of the German \ac{aFRR} market. Future works should extend the scope of use cases with diverse timescales and energy-use characteristics.

\bibliographystyle{ieeetr}
\bibliography{reference.bib}

\end{document}